# Microwave Specular Returns and Ocean Surface Roughness

Paul A. Hwang, Thomas L. Ainsworth, *Fellow, IEEE*, and Jeffrey D. Ouellette

*Abstract (250 words limit)* — Remote sensing measurements have been an important data source of ocean surface roughness. With many satellite platforms, data in extreme wind conditions from global oceans have become more abundant. Since ocean remote sensing is interdisciplinary, it requires coherent consideration from both remote sensing and oceanographic perspectives in order to access this precious ocean data source. Scatterometers operating at moderate and high incidence angles provide information on the Bragg resonance spectral components of the ocean surface waves. Monostatic and bistatic reflectometers provide spectrally integrated information of ocean waves longer than several times the incident electromagnetic (EM) wavelengths. Tilting modification of the local incidence angle for the specular facets located on slanted background surfaces is an important factor in relating the lowpass mean square slope (LPMSS) and microwave specular returns. For very high wind condition, it is necessary to consider the modification of relative permittivity by air in foam and whitecaps produced by wave breaking. This paper describes the application of these considerations to monostatic and bistatic microwave specular returns from the ocean surface. Measurements from Ku and Ka band altimeters and L band reflectometers are used for illustration. It remains a challenge to acquire sufficient number of high-wind collocated and simultaneous reference measurements for algorithm development or validation and verification effort. Solutions from accurate forward computation can supplement the sparse high wind databases. Modeled specular normalized radar cross sections (NRCSs) for L, C, X, Ku, and Ka bands with wind speeds up to 99 m/s are provided.

## I. Introduction

The range of ocean surface wavelengths important to microwave remote sensing extends several orders of magnitude. Crombie [1] reports the Doppler frequency spectrum of 13.56 MHz radar sea echo at low grazing angle. A distinct spectral peak at 0.38 Hz is illustrated, corresponding to the resonance ocean surface wavelength of about 10 m (wavenumber $k$ about 0.6 rad/m). He goes on to suggest that variable frequency equipment can be used to measure the ocean surface wave spectrum. Depending on frequency and incidence angle, the range of resonance surface wavenumbers spans from about 500 rad/m (Ku band) to about 20 rad/m (L band) for microwaves sensors operating at moderate and high incidence angles [2, 3].

For altimeters and reflectometers, the specular reflection mechanism dominates. The normalized radar cross section (NRCS) is proportional to the number of specular points and the average radii of curvature of the specular reflection facets [4-6]. With the Gaussian distribution describing the elevation and slope of the moving ocean surface [7], a simple inverse relationship between NRCS and surface mean square slope (MSS) is established [5-6]. Further analysis [8-10] indicates that the responsible MSS is contributed by surface waves longer than the EM wavelengths. The frequently cited ratio between the upper cutoff wavenumber $k_u$ of lowpass MSS integration and the EM wavenumber $k_r$ is between 3 and 6 [9-10]. Thus, for Ku band (~14 GHz) altimeter, $k_u$ is about 50 to 100 rad/m, and for L band (~1.6 GHz) reflectometer it is about 6 to 11 rad/m.

These theoretical and empirical analyses provide useful guidelines for quantitative investigation of the connection between specular NRCS and ocean surface roughness. Ku-band altimeters have been in operation for many decades and there is a rich trove of well-calibrated Ku band altimeter NRCS data for a close examination of the specular point theory applied to nadir-looking altimeters: $\sigma_0 = |R(0)|^2 / s_f^2$, where $\sigma_0$ is NRCS, $R(0)$ is the Fresnel reflection coefficient for normal incidence, and $s_f^2$ is the Ku-band LPMSS [8-10]. One peculiar outcome is that the resulting $s_f^2$ calculated from measured Ku-band NRCS is larger than the optical total MSS [11-12]. The difference is especially obvious in low and moderate wind speeds ($U_{10}$ less than about 10 m/s). To address this paradox, an effective reflectivity ranging from 0.34 to 0.5 has been suggested [8-10]; those numbers are much smaller than the nominal value of 0.62 computed from the Ku band relative permittivity. An alternative explanation is that the peculiar result can be reconciled if the tilting effect of the reflecting specular facets is considered when applying the specular point theory [13-14]. It has been about two decades since the study of [13-14] and our understanding of the ocean surface wave spectrum has advanced considerably with incorporation of remote sensing data into the relatively sparse databases of short-scale ocean surface waves accumulated from direct observations [15-17]. Here we revisit the Ku-band altimeter NRCS analyses. The results are applied to the bistatic observations of L band LPMSS [18-21] and recent reports of

Manuscript received xxxx. This work was supported by the Office of Naval Research under Grant No. N0001416WX00044.
The authors are with the U. S. Naval Research Laboratory, Washington DC 20375 USA (e-mail: paul.hwang@nrl.navy.mil; tom.ainsworth@nrl.navy.mil; Jeffrey.Ouellette @nrl.navy.mil).

Color versions of one or more of the figures in this article are available online at http://ieeexplore.ieee.org.
Digital Object Identifier 10.1109/TGRS.20xx.

bistatic NRCS results derived from the NASA CYclone Global Navigation Satellite System (CYGNSS) mission [22-24].

Sec. II gives a brief review of the specular point theory [4-6, 8]. Sec. III describes its application to monostatic and bistatic observations. Sec. IV discusses issues such as wind speed and wave age relationship on LPMSS, whitecap effects caused by wave breaking on surface reflectivity in high winds, extending the analysis to various frequencies, and other theories on specular returns such as [25-26]. Sec. V is summary.

## II. REVIEW OF SPECULAR POINT THEORY

Kodis [4] presents a theoretical analysis of backscattering from a perfectly conducting 1D irregular surface at very short EM wavelengths. The integral formulas are derived directly from the vector field theory. He shows that to the first order of approximation, the backscattering cross section is proportional to the product of the average number of specular points illuminated by the EM waves $n_A$, and the geometric mean of the two principal radii of curvature of those specular points $r_1$ and $r_2$, i.e.,

$$\sigma(k_i, -k_i) \sim \pi \langle |r_1 r_2| \rangle n_A, \quad (1)$$

where $k_i$ is the incidence EM wavenumber. This analysis elicits the close connection between EM scattering and statistical and geometrical properties of the rough surface. In order to carry out the calculations further, it is necessary to specify the statistics of the rough surface, in particular in regard to the average number of illuminated specular points and their average curvature.

Barrick [5-6] extends the analysis to 2D horizontal plane, bistatic configuration, polarization states, and finite surface conductivity. Following his notations as defined by the scattering geometry depicted in his Fig. 1, which is reproduced here as Fig. 1, the NRCS for arbitrary incident and scattered polarization states ($\eta$ and $\xi$, respectively), incidence angles ($\theta_i$, $\phi_i=0$), and scattering angles ($\theta_s$, $\phi_s$) is

$$\sigma_{0\xi\eta} = \pi |R_{\xi\eta}(\iota)|^2 \langle |r_1 r_2| \rangle n_A, \quad (2)$$

where $R_{\xi\eta}(\iota)$ is the reflection coefficient from infinite plane tangent to the surface at the specular points for incidence and scattered states, and $\iota$ is the local (effective) incidence angle at the specular points. From Kodis's stationary phase analysis [4], it is shown that the effective incidence angle $\iota$ is half the angle between the incidence and scattering propagation directions [5], and it can be expressed as a function of incidence and scattering angels: (after correcting a couple of typographic errors)

$$\cos \iota = [(1 - \sin\theta_i \sin\theta_s \cos\phi_s + \cos\theta_i \cos\theta_s)/2]^{1/2}. \quad (3)$$

The average number of specular points per unit area for a rough surface and the average radii of curvatures are then derived in terms of the surface statistics. Sec. II and Sec. III of [5] and Appendix B of [6] give the detailed mathematical derivation of $n_A$ and $\langle |r_1 r_2| \rangle$. The final results are copied here:

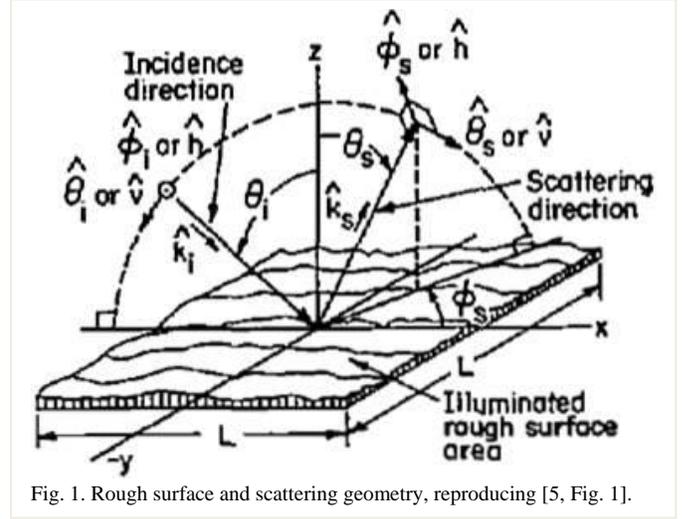

Fig. 1. Rough surface and scattering geometry, reproducing [5, Fig. 1].

$$n_A = \frac{7.255}{\pi^2 l^2} \exp\left(\frac{-\tan^2 \gamma}{s_f^2}\right), \quad (4)$$

$$\langle |r_1 r_2| \rangle = \frac{0.138 \pi l^2}{s_f^2} \sec^4 \gamma, \quad (5)$$

where $l$ is the correlation length between surface points $\zeta(x, y)$ and $\zeta(x', y')$ separated by a horizontal distance $r = [(x - x')^2 + (y - y')^2]^{1/2}$ and assuming a Gaussian distribution of the correlation coefficient as $r \to 0$. The numerical constants in (4) and (5) result from carrying out triple integral functions of ($\zeta_{xx}$, $\zeta_{xy}$, $\zeta_{yy}$); the details are provided in [5-6].

Substitute (4) and (5) into (2), then

$$\sigma_{0\xi\eta} = |R_{\xi\eta}(\iota)|^2 \frac{\sec^4 \gamma}{s_f^2} \exp\left(\frac{-\tan^2 \gamma}{s_f^2}\right), \quad (6)$$

where $s_f^2$ is the ocean surface LPMSS, and $\tan\gamma$ is the surface slope at the specular point, which can be expressed as a function of incidence and scattering angles

$$\tan\gamma = \frac{(\sin^2\theta_i - 2\sin\theta_i \sin\theta_s \cos\phi_s + \sin^2\theta_s)^{1/2}}{\cos\theta_i + \cos\theta_s}. \quad (7)$$

A case of special interest is backscattering: $\phi_s = \pi$, $\theta_s = \theta_i$, $\iota = 0$, $\gamma = \theta_i$, and the NRCS is

$$\sigma_{0\xi\eta} = |R_{\xi\eta}(0)|^2 \frac{\sec^4 \theta_i}{s_f^2} \exp\left(\frac{-\tan^2 \theta_i}{s_f^2}\right). \quad (8)$$

## III. APPLICATION

### A. Monostatic Ku band

For a nadir-looking altimeter ($\theta_i = 0$) with linear polarization ($h$ or $v$ for horizontal or vertical), (8) becomes

$$\sigma_0 = \frac{|R(0)|^2}{s_f^2} = \sigma_{0hh} = \sigma_{0vv}, \quad (9)$$

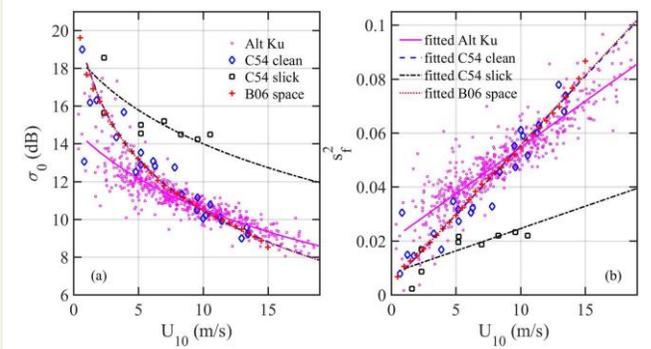

Fig. 2. (a) Ku band altimeter NRCS and comparison with computational results using (9) with the optical MSS in clean and slick waters: C54 and B06. (b) The MSS computed using (9) with the Ku band altimeter NRCS and comparison with the optical MSS in clean and slick waters. The smooth curves are linear fitting to the three sets of MSS data. The smooth curves in (a) are the NRCS computation using (9) with the linear fitted curves of the MSS data.

where $R(0)$, shorthand for $R_{\eta\eta}(0)$ with $\eta = h$ or $v$, is the Fresnel reflection coefficient for normal incidence, and the NRCS is independent on polarization states. Applying (9) to Ku band altimeter measurements, a rather peculiar result is discovered [8-10, 13-14]: that the computed Ku-band LPMSS is larger than the total optical MSS in clean water $s_\infty^2$ [11-12].

The LPMSS $s_f^2$ is an integrated surface wave property, which is defined as

$$s_f^2 = \int_0^{k_u} k^2 S(k) dk, \qquad (10)$$

where $S$ is the surface wave elevation spectrum, and $k_u$ is the upper limit of lowpass filter, which is in turn proportional to the EM wavenumber $k_r$. The ratio $k_u/k_r$ is generally given as between 3 and 6 [9-10]. When distinction of EM frequency is desired, $s_f^2$ is also given as $s_{k_u}^2$ in this paper. For example, for Ku band EM frequency $f_r = 14$ GHz and $k_u = k_r/3 = 293/3 = 98$ rad/m, the corresponding $s_f^2$ is $s_{98}^2$ for clarification. The optical EM frequency is many orders of magnitude higher than those of the microwave sensors used in ocean remote sensing, so $s_\infty^2$ is expected to be the upper bound of $s_f^2$ observed by microwave equipment.

Fig. 2(a) shows the Ku-band altimeter NRCS collected in the Gulf of Alaska and Bering Sea [14], and the calculated NRCSs based on the optical MSS from sun glitter analyses in clean and slick waters [11-12]. Fig. 2(b) shows the optical MSS and $s_f^2$ derived from the Ku-band altimeter NRCS using (9). Two sets of MSS reported in [11] are obtained from airborne sun glitter analyses in clean water surfaces (wind speed up to 14.5 m/s) and water surfaces covered with artificial slicks created by an oil mixture (wind speed up to 10.6 m/s). The optical MSS reported in [12] is also from sun glitter analysis but from a spaceborne optical sensor (wind speed up to 15 m/s); the results are essentially identical to those of the clean water condition in [11]. For the slick waters, surface waves shorter than about 30 cm are suppressed [11], therefore the two sets of MSS are $s_\infty^2$ and $s_{21}^2$. The Ku band (14 GHz) EM wavelength is about 2.1 cm, with the factor $k_u/k_r = 3$ to 6 applied, the observed LPMSS is between $s_{98}^2$ and $s_{49}^2$. It is expected that the surface roughness sensed by the Ku band altimeter to be between the optical data in clean and slick waters. This, however, is not the case for the result in low and moderate winds ($U_{10} \leq \sim 10$ m/s) as illustrated in Fig. 2(b). Especially intriguing of the observation is that the discrepancy increases toward lower wind condition for which the ocean surface is less nonlinear and simplifications made in the EM theoretical development are better justified.

As mentioned in the Introduction, many researchers resort to using an effective reflectivity much smaller than that computed from the relative permittivity [8-10], more discussion on reflectivity is deferred to Sec. IV.B. Alternative explanations have been offered [13-14, 25-26]. The mechanisms described in [25-26] will be discussed in Sec. IV.D. In [13-14], the authors stress that (8) carries the physical meaning of an exponentially attenuating contribution with respect to the tilting angle, $\theta_i = \gamma$. This becomes clear when we reexam (6), which is the source of (8). It is developed to represent the scattering for a particular reference angle $\gamma$ as clearly given in (4) and (5). That is, the left-hand side of (6) is $\sigma_{0\xi\eta}(\theta_s, \gamma)$. This important physical meaning would not have been lost if one has not replaced $\gamma = \theta_i$ on the right-hand side of the backscattering equation, i.e., (8) is restored to

$$\sigma_{0\xi\eta}(\theta_s = \theta_i, \gamma) = \\ |R_{\xi\eta}(0)|^2 \frac{\sec^4(\gamma = \theta_i)}{s_f^2} \exp\left(\frac{-\tan^2(\gamma = \theta_i)}{s_f^2}\right). \quad (11)$$

When the altimeter solution is reduced to (9), we also constrained the ocean surface to be flat and $\gamma$ is always 0, which is the main cause of the unreasonable result illustrated in Fig. 2. The alternative explanation offered in [13-14] is that (8) can be interpreted as the specular scattering pattern with respect to the incidence angle $\theta_i$. This is illustrated in a conceptual sketch in [13, Fig. 7], which is reproduced as Fig. 3 here. In the figure, the parallel horizontal lines represent the far-field EM wave fronts emitted from zenith and impinge on the ocean surface. Five scattering patterns are illustrated. Patterns 1, 4, and 5 are from three incrementally rougher patches located on background surfaces that are locally parallel to the incoming wave fronts such that the local incidence angle is not changed from the nominal incidence angle (0 in this case). The backscattering returns from the three patches are inversely proportional to the surface roughness as expected from (9). Patterns 2, 3, and 4 are from three statistically identical roughness patches and located on background surfaces with different orientations. The backscattering strengths from the three patches observed by the monostatic antenna are different although the reflecting patches have identical statistical roughness. The difference of the returns toward the nominal incidence direction (from zenith) reflects the tilting effect as described by the exponential term in (8) or (11).

Consequently, there is one more step to obtaining the backscattering NRCS accounting for all $\gamma$ values, that is,

$$\sigma_0(\theta_s = \theta_i) = \int_{-\infty}^{\infty}\int_{-\infty}^{\infty} |R(0)|^2 \cdot$$
$$\frac{\sec^4 \theta_{il}}{s_f^2} \exp\left(\frac{-\tan^2 \theta_{il}}{s_f^2}\right) \cdot \qquad (12)$$
$$p(\tan\gamma_x, \tan\gamma_y) d\tan\gamma_x d\tan\gamma_y$$

where $\theta_{il}$ is the local incidence angle satisfying the specular reflection condition (i.e., $\theta_{il} = \theta_{sl}$ on the tilting surface, $\theta_{sl}$ is the local scattering angle), $\gamma$ is the global tilting angle, $\tan\gamma = (\tan\gamma_x, \tan\gamma_y) = (\zeta_x, \zeta_y)$ is the corresponding surface slope, and $p(\tan\gamma_x, \tan\gamma_y) = p(\zeta_x, \zeta_y)$ is the probability density function (pdf) of the background (global) surfaces that tilt the specular patches. The term inside the double integrals and before the pdf function, that is essentially (8), is now interpreted as the scattering pattern, and (12) is equivalent to the tilting modulation of local incidence angle in the discussion of scatterometer returns [8] with the pdf function accounting for the off-specular contribution. For the Gaussian distribution of sea surface slopes with equal upwind and crosswind slope components, i.e., $s_{fx}^2 = s_{fy}^2 = s_f^2/2$,

$$p(\tan\gamma_x, \tan\gamma_y) = \frac{1}{\pi s_f^2} \exp\left(\frac{-\tan^2\gamma}{s_f^2}\right). \quad (13)$$

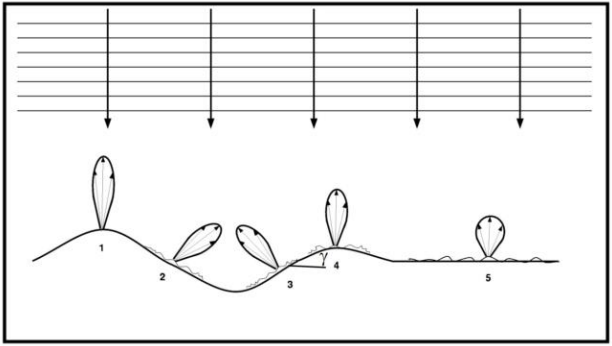

Fig. 3. A conceptual sketch depicting the scattering of radar waves by surface patches of various roughness (1, 4, and 5) and the effect of tilting background surface on the backscattering intensity from patches of identical statistical roughness (2, 3, and 4); reproducing [13, Fig. 7].

For the altimeter application, we have

$$\sigma_0 = \sigma_0(0) = \int_{-\infty}^{\infty}\int_{-\infty}^{\infty} |R(0)|^2 \cdot$$
$$\frac{\sec^4 \gamma}{s_f^2} \exp\left(\frac{-\tan^2 \gamma}{s_f^2}\right) \cdot \qquad (14)$$
$$\frac{1}{\pi s_f^2} \exp\left(\frac{-\tan^2 \gamma}{s_f^2}\right) d\tan\gamma_x d\tan\gamma_y$$

Fig. 4(a) shows the NRCS results computed with (14) and their comparison with altimeter observations. Fig. 4(b) shows the two sets of $s_f^2$ used in the computation. They are based on the H18 spectrum model ($s_{H18}^2$) [15-16] integrated to $k_u = k_r/3$ and $k_r/5$. The agreement between the theoretical computation and measurement is improved considerably. More discussion on LPMSS analysis and reflectivity $|R(0)|^2$ is deferred to Sec. IV.

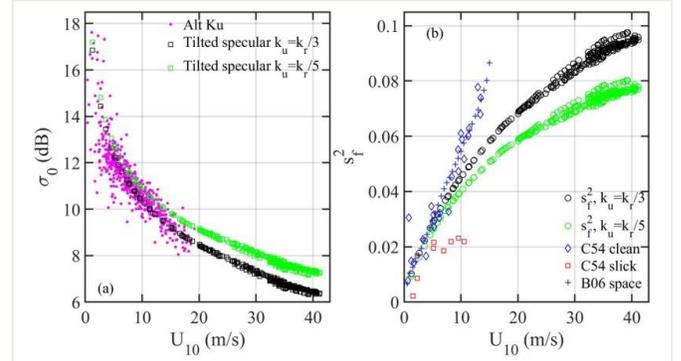

Fig. 4. (a) Ku band NRCS and comparison with model computation (14). (b) The LPMSS used for the NRCS computation shown in (a), and comparison with the optical MSS in clean and slick waters. The LPMSS is computed with the H18 spectrum model integrated to $k_r/3$ and $k_r/5$.

*B. Bistatic L band*

Tilting modification is expected to impact both bistatic and monostatic specular reflections. Applying the same procedure discussed in the monostatic case to the bistatic solution (6) for $\sigma_{0\xi\eta}(\theta_s, \gamma)$, the BNRCS for 2D Gaussian pdf of tilting surfaces is:

$$\sigma_0(\theta_s) = \int_{-\infty}^{\infty}\int_{-\infty}^{\infty} |R_{\xi\eta}(\iota)|^2 \frac{\sec^4\gamma}{s_f^2} \exp\left(\frac{-\tan^2\gamma}{s_f^2}\right) \cdot$$
$$\frac{1}{\pi s_f^2} \exp\left(\frac{-\tan^2\gamma}{s_f^2}\right) d\tan\gamma_x d\tan\gamma_y \qquad . (15)$$

Consider the simpler case of scattering in the same transmission plane (Fig. 5), there are three possibilities: (a) $\theta_i > 2\gamma$, (b) $\gamma < \theta_i \leq 2\gamma$, and (c) $\theta_i < \gamma$ (and their mirror images). For case (a) $\phi_s = \pi$ and $\theta_s = \theta_i - 2\gamma$, and for cases (b) and (c) $\phi_s = 0$ and $\theta_s = 2\gamma - \theta_i$. Substituting these scattering angles into (3), we get

$$\cos\iota = \left[(1+\cos 2\gamma)/2\right]^{1/2}. \qquad (16)$$

For ocean applications $\gamma$ is small. For example, the L band $s_f^2$ is generally less than 0.05 even in tropical cyclone (TC) conditions (for Ku band, it's less than about 0.1, see Sec. IV.A). The representative value of $\gamma$ is $\tan^{-1}s_f$, which is less than 0.22 (L band) to 0.31 (Ku band) even in TCs. For non-TC conditions, the corresponding values are much smaller. With perturbation approximation, $\cos 2\gamma \approx 1 - (2\gamma)^2/2$, then $\iota \approx \gamma$ for small $\gamma$.

It turns out that for global positioning system reflectometry (GPSR) with right-hand-circular transmit and left-hand-circular receive, the $|R_{lr}(\iota)|^2$ is almost independent on $\iota$ up to about 60° [15] so $|R_{lr}(\iota)|^2$ plays a rather minor role for computing circular polarization BNRCS except at low grazing angles. The reflectivity in (6) or (15) can be approximated by $|R_{lr}(0)|^2$ for GPSR from the ocean surface.

Through GPSR delay Doppler waveform analyses there are now several sets of L band LPMSS collected in TC conditions [18-21]. These data are identified as $s^2_{GPSR}$ and illustrated in

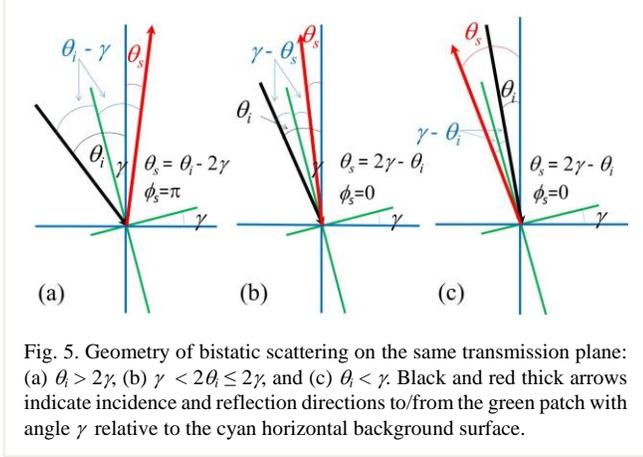

Fig. 5. Geometry of bistatic scattering on the same transmission plane: (a) $\theta_i > 2\gamma$, (b) $\gamma < 2\theta_i \leq 2\gamma$, and (c) $\theta_i < \gamma$. Black and red thick arrows indicate incidence and reflection directions to/from the green patch with angle $\gamma$ relative to the cyan horizontal background surface.

Fig. 6 (labeled K0913 [18-19] and G1318 [20-21] in the legend); the least-squares fitted curve is given by the solid black line (labeled KG). The $s^2_{GPSR}$ have served to address one of the most unsettled issues in the ocean surface wind wave spectrum function $S(f)$, i.e., the spectral slope in the high frequency region [27-28]. The clarification of spectral slope is especially critical to the LPMSS determination using an ocean wave spectrum model. The refinement of the $S(f)$ function, in turn, offers the feasibility to derive LPMSS given $U_{10}$ and windsea dominant wave period $T_p$ from operational-system measurements [16] or to create synthetic high-wind LPMSS with a small number of critical TC parameters [15, 28]. Also shown in Fig. 6 are the LPMSS computed from the G18 ($s^2_{G18}$) [28] and H18 ($s^2_{H18}$) [15] spectrum models with $k_u = k_r/3$ and $k_r/5$. In addition, the results based on E97 ($s^2_{E97}$) [29] are also displayed; the E97 is used in the CYGNSS project [30-31]. For comparison, the optical data obtained in clean and slick sea surfaces [11-12] are illustrated with black markers in the figure. The GPSR data have expanded the wind speed coverage considerably (from 15 to 59 m/s), and they are critical for refining the ocean surface wind wave spectrum models in high wind conditions. More detail on deriving LPMSS from a wave spectrum is deferred to Sec. IV.A.

Retrieving BNRCS from GPSR, or global navigation satellite system reflectometry (GNSSR) in general, remains a challenging task [22-24]. Here we investigate the properties of L-band BNRCS through forward computation using (15), together with the input of L-band LPMSS such as those illustrated in Fig. 6. The results are then compared with the most recent publications of L band BNRCS results from the CYGNSS mission [23-24].

Fig. 7 shows the results computed with (15) and the LPMSS derived from the three spectrum models shown in Fig. 6. Superimposed with the computation curves are two sets of CYGNSS BNRCS results: the black crosses and pluses labeled R18 [23] are the GMF at 10° and 50°, and the black squares labeled B20 [24] are the incidence-angle-averaged result. It is clear that the CYGNSS BNRCS processing is still evolving. Interestingly, there is a good agreement between the most recent version [24] (B20: black squares) and the computed BNRCS with $s^2_{G18}$ and $s^2_{H18}$ integrated to $k_u=k_r/3$; the model-data difference is less than about 1 dB to wind speed up to about

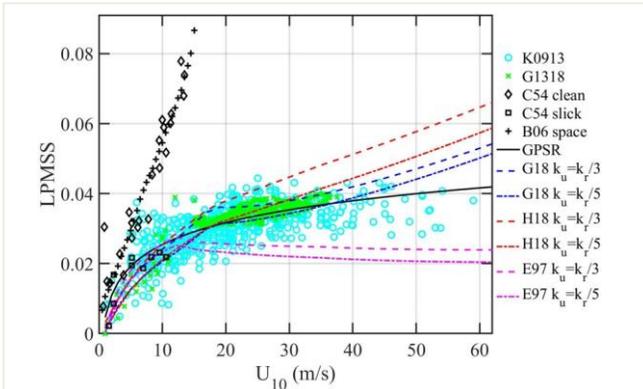

Fig. 6. L band LPMSS derived from GPSR delay Doppler analyses (labeled K0913 and G1318), the least-squares curve fitted through all the GPSR data is given by the solid black line (labeled KG). Three sets of computation from spectrum models (E97, G18 and H18) with $k_u = k_r/3$ and $k_r/5$ are also illustrated. For comparison, optical MSS in clean and slick waters are shown with black markers.

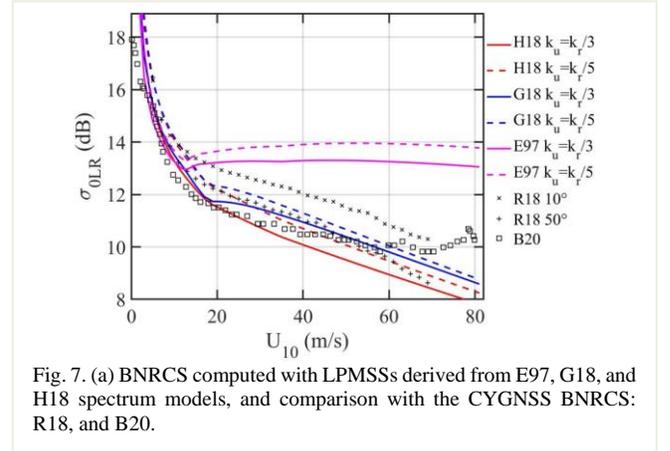

Fig. 7. (a) BNRCS computed with LPMSSs derived from E97, G18, and H18 spectrum models, and comparison with the CYGNSS BNRCS: R18, and B20.

60 m/s. The computed BNRCS with $s^2_{E97}$ fares far worse because E97 underestimates the LPMSS for $U_{10}$ greater than about 14 m/s (Fig. 6). It illustrates the importance of choosing an accurate surface wave spectrum model for ocean remote sensing analysis. More discussion on LPMSS and reflectivity $|R_{\zeta\eta}(\iota)|^2$ is deferred to Sec. IV.

## IV. DISCUSSION

### A. LPMSS, wind speed, and wave development stage

The functional form of the ocean surface wind wave spectrum remains one of the most uncertain quantities in ocean remote sensing problems. For specular return, the relevant property is the LPMSS with the upper integration wavenumber $k_u$ determined by the EM wavenumber $k_r$. From altimeter analyses [9-10] the range of $k_u/k_r$ is generally determined to be between 3 and 6. In this paper, we have shown results of specular computation obtained with $k_u/k_r = 3$ and 5, which corresponds to $k_u = 11$ and 6.6 rad/m for L band (1.575 GHz), and 98 and 59 rad/m for Ku band (14 GHz). The contribution of long surface waves in the energetic dominant wave region becomes more important as wind speed increases and EM frequency decreases.

The value of the wave spectral slope –$s$ in the high frequency region of the wind wave frequency spectral function is one of the most uncertain spectral properties critical to the determination of LPMSS. Traditional wind wave spectrum models assume $s$ to be either 4 or 5 [32-35], although field observations have shown a wide range between about 2 and 7 [27, 36]. The wave spectral level drops sharply toward both high and low frequencies from the spectral peak, therefore, the wave height measurements are not very sensitive for addressing the spectral slope issue. Wave slope data are much more useful for this task [15-16, 27-28].

For many decades, the airborne sun glitter analyses in clean and slick waters reported in 1954 [11] have remained the most comprehensive ocean surface MSS dataset, the wind speed range is between 0.7 and 13.5 m/s for the clean water condition, and between 1.6 and 10.6 m/s for the slick water condition. The spaceborne sun glitter analysis reported in 2006 [12] expands slightly the wind speed range of clean water condition to about 15 m/s and the results are essentially identical to those of the clean water condition reported in 1954 [11]. The recent results of $s^2_{GPSR}$ further extend the wind speed coverage to 59 m/s [18-21]. With the EM frequency of 1.575 GHz, the $k_u$ is between about 5 and 11 rad/m and $s^2_{GPSR}$ data are most useful for investigating the wind wave spectrum slope. The study leads to establishing a general wind wave spectrum function G18 [28], the applicable upper limit ($k_{max}$) of the G18 spectrum function is estimated to be about the upper range of L band $k_u$ (11 rad/m). For Ku band application, the hybrid model H18 is more suitable [15-16, 28]. The H18 model uses G18 for long waves and H15 [37] for short waves, with linear matching between $k = 1$ and $4$ rad/m; the detail is described in [15].

Wind speed $U_{10}$ and windsea dominant wave period $T_p$ are the only required input for computing the G18 and H18 spectrum (and many other spectrum models such as the E97 discussed in Sec. III). The combination of $U_{10}$ and $T_p$ can be expressed as the dimensionless spectral peak frequency $\omega_\# = U_{10}/c_p = U_{10}/(gT_p/2\pi)$, where $c_p$ is the wave phase speed of the spectral peak component, and $g$ is gravitational acceleration. The inverse of $\omega_\#$ is wave age, which represents the stage of wave development. Determining the wave spectrum requires consideration of both wind speed and wave development stage. With a wave spectrum function, the $s^2_{k_u}$ can be pre-calculated for a range of $U_{10}$, $\omega_\#$, and $k_u$. For example, Fig. 8(a) and 8(b) show the contour maps of H18 $s^2_{98}$ and $s^2_{11}$, respectively. They are illustrated for $U_{10}$ between 0 and 70 m/s, and $\omega_\#$ between 0.8 and 5.2. These pre-calculated results serve as design curves (or lookup tables) for quickly obtaining the desired $s^2_f$ through interpolation. Superimposed in the figures are the observed $\omega_\#(U_{10})$ in TC and non-TC conditions. Because the wave age $1/\omega_\#$ is defined as the ratio of dominant wave phase speed $c_p$ and wind speed $U_{10}$, $U_{10}$ and $\omega_\#=U_{10}/c_p$ are not independent variables. The observed $\omega_\#(U_{10})$ in TC and non-TC conditions show the general linear relationship with a narrow range of $\omega_\#$ variation for a particular wind speed $U_{10}$ as illustrated by the color markers in Fig. 8(a) and 8(b).

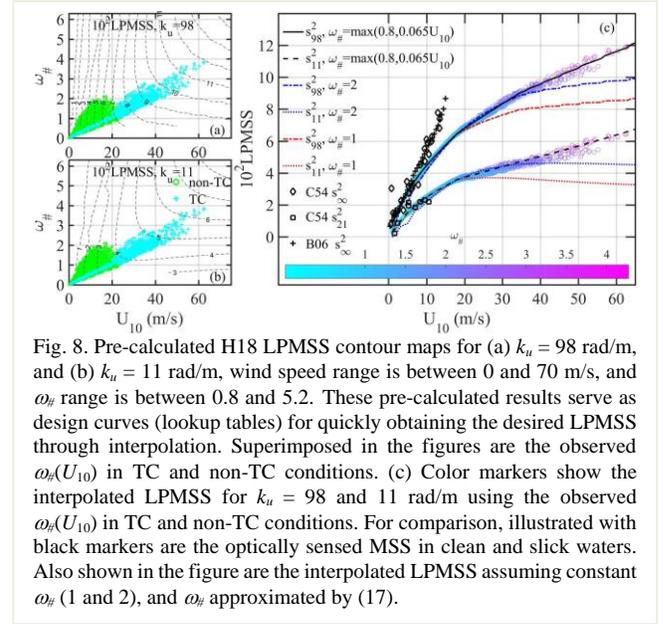

Fig. 8. Pre-calculated H18 LPMSS contour maps for (a) $k_u = 98$ rad/m, and (b) $k_u = 11$ rad/m, wind speed range is between 0 and 70 m/s, and $\omega_\#$ range is between 0.8 and 5.2. These pre-calculated results serve as design curves (lookup tables) for quickly obtaining the desired LPMSS through interpolation. Superimposed in the figures are the observed $\omega_\#(U_{10})$ in TC and non-TC conditions. (c) Color markers show the interpolated LPMSS for $k_u = 98$ and 11 rad/m using the observed $\omega_\#(U_{10})$ in TC and non-TC conditions. For comparison, illustrated with black markers are the optically sensed MSS in clean and slick waters. Also shown in the figure are the interpolated LPMSS assuming constant $\omega_\#$ (1 and 2), and $\omega_\#$ approximated by (17).

The interpolated H18 $s^2_{98}$ and $s^2_{11}$ are presented with color markers in Fig. 8(c). For comparison, the optically sensed $s^2_\infty$ in clean water [11-12] and $s^2_{21}$ in slick water [11] are illustrated with black markers in the figure. The procedure to compute $s^2_f$ with $U_{10}$ and $T_p$ input can be applied to other spectrum models, for example, the L band LPMSSs computed with the E97, G18, and H18 spectrum models have been discussed in Fig. 5.

Also shown in Fig. 8(c) are the interpolated H18 $s^2_{98}$ and $s^2_{11}$ assuming constant $\omega_\#$ (1 and 2 are used for illustration). As wind speed increases, the difference increases between $s^2_f$ computed with constant and observed $\omega_\#$. Interestingly, if the approximation

$$\omega_\# = \max\left(0.8, 0.065 U_{10}\right) \qquad (17)$$

is employed, the resulting $s^2_f$ is very close to that obtained with observed $\omega_\#$. For the purpose of obtaining $s^2_f$ from a wave spectrum function, approximation (17) simplifies the procedure in practical applications since it requires only the $U_{10}$ input (with $k_u$ specified).

The H15 and H18 spectrum functions have been used for forward simulations of L, C, and Ku scatterometer NRCS. The results are in good agreement with the respective GMFs to within about 1 to 2 dB with wind speeds in calm to TC conditions, as summarized in [38]. They are also used in forward simulations of microwave brightness temperature in L, C, X, Ku, K, and Ka bands (1.4 to 37 GHz) [39-40]. The results are in good agreement with measurements in calm to TC conditions [41-49]. In this study, we further expand the model-measurement comparison to monostatic and bistatic specular returns (Sec. III, and further discussion in Sec. IV.C), the comparison results are also quite good.

## B. Surface reflectivity considering wave breaking

The reflectivity $|R_{\xi\eta}(t)|^2$ is a function of relative permittivity and generally treated as a constant for a given EM frequency (with the assumption of some representative sea surface temperature and sea surface salinity; 293 K and 35 psu are used throughout this paper). In high winds when air is entrained by wave breaking into the water surface layer and foam covers the water surface, the modification of relative permittivity by the mixed air needs to be considered. Through analyses of microwave radiometer measurements collected in TCs that cover a wide range of frequencies, incidence angles, and both horizontal and vertical polarizations [41-49], the effects of surface foam generated by wave breaking are expressed as a function of wind speed, microwave frequency, and incidence angle [39-40]. The effective air fraction $F_a$ is related to the whitecap coverage $W_c$ as described in Appendices A and B in [40]. A brief summary is given here. The proposed $F_a/W_c$ function is

$$\frac{F_a}{W_c} = \max\left[1, \left(\frac{f}{f_{ref}}\cos^\alpha \theta\right)^\beta\right]. \quad (18)$$

From empirical fitting the computed brightness temperature with observations [41-49], the following values are recommended for the three parameters in (18):
$f_{ref}$ =14 GHz,
$\alpha$=1.3, and

$$\beta = \max\left\{0, 0.5 - \min\left\{0.5, 0.5\left[\exp\left(1.1f/f_{ref}\right) - 1.5\right]\right\}\right\}.$$

The whitecap fraction $W_c(U_{10})$ is defined by

$$W_c = \begin{cases} 0, & u_* \leq 0.11\,\text{m/s} \\ 0.30(u_* - 0.11)^3, & 0.11 < u_* \leq 0.40\,\text{m/s} \\ 0.07u_*^{2.5}, & u_* > 0.40\,\text{m/s} \end{cases}, \quad (19)$$

where $u_*$ is the wind friction velocity. The formula for the drag coefficient to obtain $u_*$ from $U_{10}$ input is given by

$C_{10} =$
$$\begin{cases} 10^{-4}\left(-0.0160U_{10}^2 + 0.967U_{10} + 8.058\right), & U_{10} \leq 35\,\text{m/s} \quad (20) \\ 2.23\times 10^{-3}\left(U_{10}/35\right)^{-1}, & U_{10} > 35\,\text{m/s} \end{cases}$$

With the effective air fraction determined from the whitecap observations, the effective relative permittivity $\varepsilon_e$ is computed with the refractive mixing rule [50-52]:

$$\varepsilon_e = \left[F_a\varepsilon_a^{1/2} + (1-F_a)\varepsilon_{sw}^{1/2}\right]^2, \quad (21)$$

where $\varepsilon_a$ and $\varepsilon_{sw}$ are the relative permittivities of air and (foamless) sea water, respectively. The effective relative permittivity $\varepsilon_e$ is then used to compute the Fresnel reflection coefficient with wind speed dependence as a result of foam effects caused by surface wave breaking.

Fig. 9 shows the Fresnel reflectivity at 0°, 10°, 30°, and 50° incidence angles for Ku (14 GHz) and L (1.575 GHz) frequencies. The Ku band altimeters discussed in this paper operate with linear polarizations (*h* and *v*), so $|R_{hh}(\theta)|^2$ and $|R_{vv}(\theta)|^2$ are illustrated. The GPSR signals are right-hand-circular transmit and left-hand-circular receive, so for L band $|R_{lr}(t)|^2$ is given, its dependence on incidence angle is very weak up to about 60° incidence angle [15]. For specular reflection from the ocean surface, $|R_{\xi\eta}(0)|^2$ is the quantity of interest (Sec. III.B) and it is independent on polarization states *hh*, *vv*, and *lr*, but can vary considerably with wind speed as a result of air entrainment by wave breaking. The foam modification is more severe toward higher frequency as expected; this can be seen from comparing the black solid line of L band and red/green (overlapped) solid lines of Ku band.

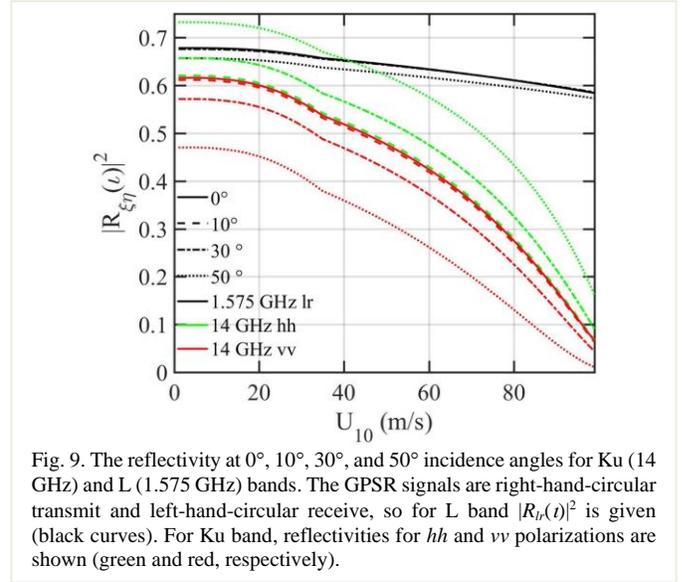

Fig. 9. The reflectivity at 0°, 10°, 30°, and 50° incidence angles for Ku (14 GHz) and L (1.575 GHz) bands. The GPSR signals are right-hand-circular transmit and left-hand-circular receive, so for L band $|R_{lr}(t)|^2$ is given (black curves). For Ku band, reflectivities for *hh* and *vv* polarizations are shown (green and red, respectively).

## C. Frequency dependence and verification

With a surface wave spectrum, the LPMSS can be computed for application to any EM frequencies. For example, Fig. 10 shows $s_{H18}^2$ integrated to $k_u = k_r/3$ (black curves) and $k_u = k_r/5$ (red curves) for several microwave frequency bands used frequently in ocean remote sensing (L, C, X, Ku, and Ka) with wind speeds up to 99 m/s. Also illustrated for comparison are the optical MSS, which is restricted to wind speed below about 15 m/s [11-12]. The specular NRCS can be computed with the LPMSS input, as illustrated in Fig. 11. For clarity, only results based on $s_{H18}^2$ integrated to $k_u = k_r/3$ are shown. It is of interest to verify the model computations with available data.

The Ku band altimeter systems have many more well-calibrated NRCS datasets compared to other frequencies due to the long history of using Ku band for ocean wind sensing, e.g., [9-10, 13-14, 53-65]. Three examples are examined here:

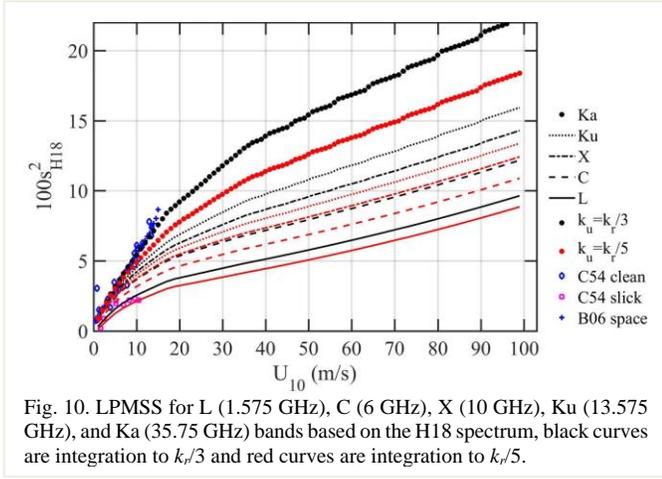

Fig. 10. LPMSS for L (1.575 GHz), C (6 GHz), X (10 GHz), Ku (13.575 GHz), and Ka (35.75 GHz) bands based on the H18 spectrum, black curves are integration to $k_r/3$ and red curves are integration to $k_r/5$.

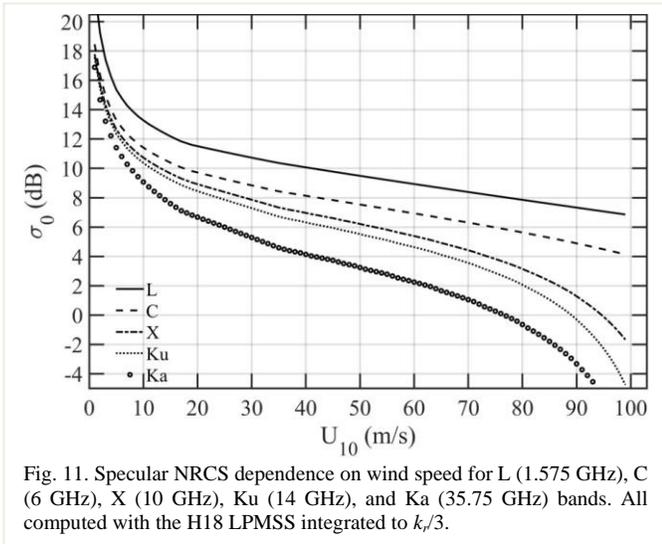

Fig. 11. Specular NRCS dependence on wind speed for L (1.575 GHz), C (6 GHz), X (10 GHz), Ku (14 GHz), and Ka (35.75 GHz) bands. All computed with the H18 LPMSS integrated to $k_r/3$.

(a) TOPEX/POSEIDON (T/P) altimeter NRCS (13.575 GHz) and collocated National Oceanographic and Atmospheric Administration (NOAA) National Data Buoy Center (NDBC) buoy datasets in three geographic regions (Gulf of Alaska and Bering Sea, Gulf of Mexico, and Hawaii islands) have been reported in [13-14] with up to 7 years of simultaneous measurements and a total of 2174 ($U_{10}$, $\sigma_0$) pairs. The maximum wind speed in the datasets is about 20 m/s. The maximum temporal and spatial differences between buoy and altimeter data are 0.5 h and 100 km, respectively; detailed information on the processing and merging of buoy and altimeter datasets is given in [13]. These data are referred to as H02 from here on.

(b) A one-year Tropical Rainfall Mapping Mission (TRMM) Precipitation Radar (TPR, 13.8 GHz) dataset is reported in [10] with more than $1.13 \times 10^7$ ($U_{10}$, $\sigma_0$) pairs. The TPR dataset is quite unusual because the wind sensor and altimeter are on the same satellite. The nadir footprint of the TRMM Microwave Imager (TMI) wind sensor is collocated with the footprint of the TPR and there is no need for temporal interpolation. The spatial resolution of TPR altimeter is about 4.3 km and that of the TMI is about 25 km, so the spatial separation between altimeter NRCS and TMI wind speed data is no more than ±12.5 km. The maximum wind speed in the dataset as presented in their Fig. 4 is about 29 m/s. These data are referred to as F03 from here on.

(c) An extensive collection of 33 years wind speed, wave height, and altimeter NRCS from 13 satellite missions ranging from Geosat to Sentinel-3A and Jason-3 is reported in [65]. The altimeter NRCS is merged with European Centre for Medium-Range Weather Forecasts (ECMWF) model wind speed in 1°×1° grids. Fig. 3a in [65] shows an example of the Jason-3 (J3) NRCS dependence on wind speed. The maximum wind speed is about 20 m/s. These data are referred to as R19 from here on.

Fig. 12 shows the bin-averaged Ku band altimeter data from the three examples described above, the data are labeled H02, F03, and R19. The H02 T/P results (red pluses) are processed from our in-house data sets. The R19 J3 results (red circles) are digitized from [65, Fig. 3a]. The F03 TPR results (blue diamonds) are digitized from [10, Fig. 4]. Compared to the T/P and J3 results, there is a 1-dB systematic bias in the TPR data, which is subtracted in the figure. A 1.92 dB bias is reported in [10] from comparing with the modified Chelton and Wentz (MCW) GMF [55, Table 1]. The MCW is designed for the Geosat altimeter. With improved algorithm and including atmospheric correction, the T/P NRCS differs from Geosat NRCS by 0.7 dB [56]. The MCW GMF with 0.7-dB adjustment is shown with the red solid line, which goes through the center of T/P, J3, and adjusted TPR data. The blue dashed line is the TPR GMF with 1 dB adjustment applied. The black line is the model computation with $s^2_{H18}$ integrated to $k_u = k_r/3$. The agreement is very good between model computation and all three Ku band data sets. In high winds, the model computation

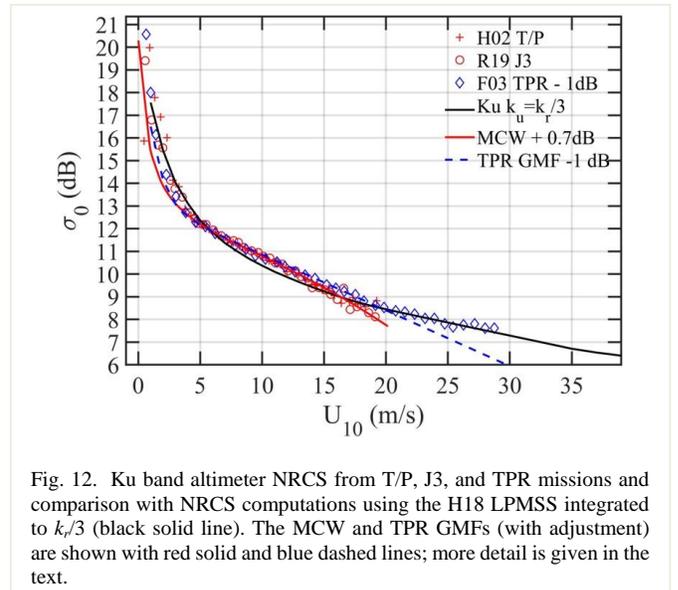

Fig. 12. Ku band altimeter NRCS from T/P, J3, and TPR missions and comparison with NRCS computations using the H18 LPMSS integrated to $k_r/3$ (black solid line). The MCW and TPR GMFs (with adjustment) are shown with red solid and blue dashed lines; more detail is given in the text.

even outperforms empirically established GMFs.

The number of reported data sets is considerably less in other EM frequencies. As mentioned in Sec. III, there are two L band BNRCS data sets reported from the CYGNSS mission [23-24]. A couple of Ka band altimeter (AltiKa) data sets have also been reported from the SARAL (the Satellite with ARgos and ALtiKa) mission [66-67]. In [66], the authors analyze all available data from AltiKa cycle 3 (23 May to 27 June 2013). Coincident ECMWF wind speeds are the wind reference. The

> LspecularBNRCS< 9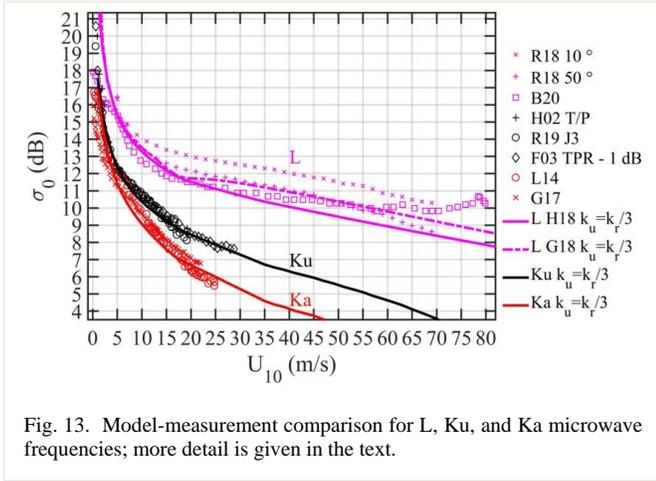

Fig. 13. Model-measurement comparison for L, Ku, and Ka microwave frequencies; more detail is given in the text.

distribution of NRCS and $U_{10}$ data points and the bin average are given in their Fig. 3a. In [67], the authors analyze the AltiKa data of cycles 4 to 14 (27 June 2013 to 17 July 2014). The wind reference is also the ECMWF wind product. The average results are given in their Fig. 3.

Fig. 13 shows the data from all three EM frequencies together and the comparison with model computation with $s^2_{H18}$ integrated to $k_u = k_r/3$. For the Ka band the model-data agreement is close to that of the Ku band comparison. For L band, the model results with both $s^2_{H18}$ and $s^2_{G18}$ integrated to $k_u = k_r/3$ are shown (solid and dashed-dotted magenta lines). They are in good agreement with measurements, especially with the most recent BNRCS report (B20 [24]). Less than 1 dB difference is found between model and measurements to wind speed about 60 m/s.

*D. Other theories and concluding remarks*

Many theories had been advanced to explain the altimeter specular scattering without compromising the Fresnel reflectivity. For example, Voronovich and Zavorotny [25] emphasize the importance of the refraction effects from waves shorter than the EM wavelengths. Their analysis shows that the coherent reflection from a flat surface with high-frequency roughness is attenuated by a factor $\exp(-2R_a^2)$, where $R_a = k_r \eta_{rmsHF}$, and $\eta_{rmsHF}$ is the RMS elevation of high frequency waves. Doing spectrum integration from $k_u$ to $\infty$, the factor $\exp(-2R_a^2)$ is not negligible, and the whole "paradox" may stem from neglecting a high-frequency part of the spectrum. Another theory is given by Nouguier et al. [26]. They emphasize the importance of nonlinearity and propose an altimeter solution that includes the fourth-order (curvature) statistics of the ocean surface, which is poorly understood for moderate and high wind conditions. The only ocean data providing the curvature statistics are from [11-12]; the maximum wind speed is 15 m/s for $s^2_\infty$ and 10.6 m/s for $s^2_{21}$. We have found out that very unrealistic results are produced when extending the equations of third- and fourth-order statistics given in [11-12] to high winds.

As commented early, one of the most interesting features in the altimeter analysis (Sec. III.A) and especially intriguing is the observation that the discrepancy between solution (9) and NRCS data increases toward low wind condition for which the ocean surface is less nonlinear and simplifications made in the EM theoretical development are better justified. Incorporating the refraction component from high frequency waves or nonlinearity consideration will be more useful for high wind conditions and less effective in addressing the discrepancy in low winds. The main reason these alternative theories also work well is that the exponential term in (8) is retained in their solutions.

An example of higher order nonlinear terms contributing little to the altimeter return at low wind is given in Brown [68, Fig. 3] illustrating the first order and second order solutions for Ku band backscattering at 4.3 m/s. The figure is reproduced as Fig. 14(a) here. At normal incidence, the second order solution (dashed-dotted line) is about 17 dB below the first order solution (dashed line), and the sum (solid line) of the two terms is indistinguishable from the first order solution (dashed line). Another example of the importance of retaining the exponential term in the altimeter solution is presented in Hwang et al. [69, Fig. 1] that shows the two-scale solution of backscattering at normal incidence and its comparison with T/P data, the figure is reproduced as Fig. 14(b) here. For small incidence angles (nominal $\theta_i < \sim 10°$), the backscattering solution (8) is subjected to tilting modulation. Because the exponential term is kept in the solution process, the results at $\theta_i = 0°$ are in good agreement with T/P observations.

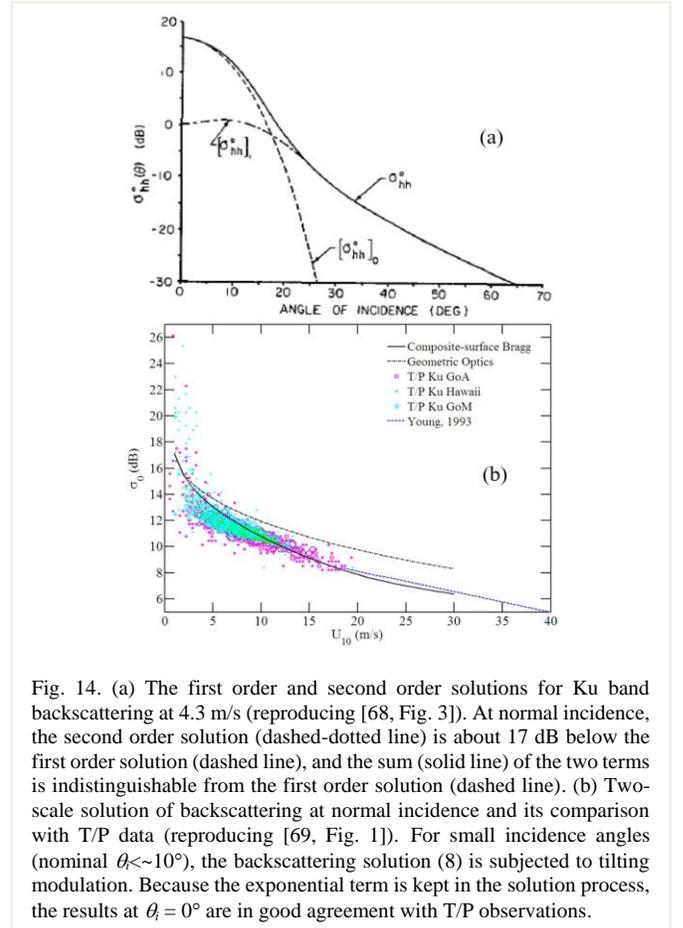

Fig. 14. (a) The first order and second order solutions for Ku band backscattering at 4.3 m/s (reproducing [68, Fig. 3]). At normal incidence, the second order solution (dashed-dotted line) is about 17 dB below the first order solution (dashed line), and the sum (solid line) of the two terms is indistinguishable from the first order solution (dashed line). (b) Two-scale solution of backscattering at normal incidence and its comparison with T/P data (reproducing [69, Fig. 1]). For small incidence angles (nominal $\theta_i < \sim 10°$), the backscattering solution (8) is subjected to tilting modulation. Because the exponential term is kept in the solution process, the results at $\theta_i = 0°$ are in good agreement with T/P observations.

As discussed in Sec. III.A, the solutions of the number of specular points per unit area (4) and the average radii of curvature of the specular reflection facets (5) are for a given $\gamma$. The closed-form solution of bistatic specular scattering (6) is $\sigma_{0\xi\eta}(\theta_s,\gamma)$ and the closed-form solution of monostatic backscattering (8) or (11) is $\sigma_{0\xi\eta}(\theta_s=\theta_i,\gamma)$. One more step is needed to obtain the monostatic and bistatic NRCS by accounting for all $\gamma$ values, which leads to (14) and (15), respectively. Specular NRCS data from a wide range of sources covering L, Ku, and Ka bands are assembled and they are all in very good agreement (Fig. 13) with the model computation using (14) and (15).

## V. Summary

The specular point theory [4-6, 8] establishes a firm relationship between specular NRCS and surface wave statistical and geometric properties in a close-form expression for a given $\gamma$ (6). Specifically, it states that the NRCS is linearly proportional to the reflectivity $|R_{\xi\eta}(t)|^2$, inversely proportional to the LPMSS $s_f^2$, and multiplied with a term dominated by the exponential attenuation with respect to the surface slope at the specular point (tan$\gamma$). The closed-form solution is developed to represent the scattering for a particular $\gamma$, that is, the left-hand side of (6) is $\sigma_{0\xi\eta}(\theta_s,\gamma)$. It is necessary to account for all $\gamma$ values for the final NRCS solution. This step is achieved by a 2D Gaussian integration, which leads to (14) for altimeters and (15) for bistatic reflectometers. The model results are in very good agreement with a wide collection of specular NRCS data from L, Ku, and Ka instruments (Fig. 13).

We reemphasize that remote sensing measurements are indisputably among the most important data sources of ocean surface roughness. The satellite platforms allow data collection in extreme wind conditions from global oceans. Such capability is unimaginable for traditional in-situ ocean sensors. Since ocean remote sensing is interdisciplinary, it requires coherent consideration from both remote sensing and oceanographic perspectives in order to access this precious ocean data source for improving the ocean surface wave spectrum that is important to remote sensing analysis. In return, the improved ocean surface wave spectrum provides more accurate forward computations of microwave altimeters, reflectometers, scatterometers, and radiometers covering a wide range of frequencies, incidence angles, and polarizations.

ACKNOWLEDGMENTS

This is U. S. Naval Research Laboratory Publication JA/7260—20-0999.